\documentclass[12pt, preprint]{aastex}
\usepackage{emulateapj5}
\newcommand\ea{et al.\ }
\newcommand\hst{{\it HST}}
\newcommand\ha{H$\alpha$}
\def\o3{[\ion{O}{3}]}
\def\s2{[\ion{S}{2}]}
\newcommand\kms{\ifmmode{\rm\,km\,s^{-1}}\else{${\rm\,km\,s^{-1}}$}\fi}
\newcommand\psc{\ifmmode{\rm\,cm^{-2}}\else{${\rm\,cm^{-2}}$}\fi}
\slugcomment{To appear in The Astrophysical Journal, October 10, 2001}
\shorttitle{Shocking Clouds}
\shortauthors{Levenson and Graham}

\begin{document}
\title{Shocking Clouds in the Cygnus Loop$^1$}
\altaffiltext{1}{Based on observations
with the NASA/ESA {\it Hubble Space Telescope}, which is
operated by the Association of Universities for Research in Astronomy, Inc.,
under NASA contract NAS 5-2655.}
\author{N. A. Levenson}
\affil{Department of Physics and Astronomy, Bloomberg Center, Johns Hopkins University, Baltimore, MD 21218}
\email{levenson@pha.jhu.edu}
\and
\author{James R. Graham}
\affil{Department of Astronomy, University of California, Berkeley, CA 94720}
\email{jrg@astron.berkeley.edu}

\begin{abstract}
With {\it Hubble Space Telescope} Wide-Field Planetary Camera 2
observations of 
the Cygnus Loop supernova remnant, we examine the
interaction of an interstellar cloud with the blast wave
on physical scales of $10^{15}$ cm.
The shock front is distorted, revealing both edge-on and
face-on views of filaments and diffuse emission,
similar to those observed on larger scales at lower resolution.
We identify individual shocks in the cloud of density
$n \approx 15 {\rm \, cm^{-3}}$ having velocity
$v_s \approx 170 \kms$.  We also find the morphologically
unusual diffuse Balmer-dominated emission of faster shocks
in a lower-density region.  The obstacle diffracts these shocks, so they
propagate at oblique angles with respect to the primary blast wave.
The intricate network of diffuse and filamentary \ha{} emission arises 
during the early stage of 
interaction between the cloud and blast wave,
demonstrating that 
complex shock propagation and emission morphology
occur before the onset of  
instabilities that destroy clouds completely.
\end{abstract}

\section{The Southeast Knot of the Cygnus Loop}
Supernova remnants and the interstellar medium act upon each other reciprocally.
The energy of supernova remnants (SNRs) heats and ionizes the interstellar
medium (ISM), and their blast waves govern mass exchange between the hot,
warm, and cool phases of the ISM.  In turn, the extant ISM determines the
evolution of SNR blast waves as they propagate through the environment
it provides.  Thus, in order to understand the large-scale structure of
the ISM, we must also discern the nature of shock evolution in inhomogeneous
media.  

The Cygnus Loop supernova remnant provides an ideal laboratory
in which to examine these issues.  It is relatively unobscured,
with $E(B-V)=0.08$ \citep{Par67}, and nearby 
($440^{+130}_{-100}$ pc; \citealt{Bla99}), so 
1\arcsec{} corresponds to $7\times10^{15}$ cm.
It is a middle-aged supernova remnant, not only in terms of absolute lifetime,
$\tau = 8000$ yr (\citealt{Lev98}, scaling for the revised distance), 
but more importantly because the interaction of the blast wave with the ISM
dominates its evolution and its appearance at all wavelengths.
At optical wavelengths, the outstanding features are due to decelerated shocks that propagate
through dense interstellar clouds.  Shocks that are reflected off the cloud
surfaces propagate back through previously shocked material, further
heating and compressing it, enhancing X-ray emission \citep{Hes86}. 

One example of this characteristic interaction is the southeast knot.
\citet*{Fes92} drew attention to this apparently insignificant feature in
the optical and X-ray emission, which has a small angular extent compared to
the diameter of the Cygnus Loop.
In contrast to the SNR as a whole,
the optical appearance of the southeast knot 
suggests that it represents only a very minor
enhancement in the local ISM, or that the interaction is very young.
\citet{Fes92} support the former interpretation,
proposing that this 
represents the late stage of interaction with a small cloud
that has been completely engulfed by
the blast wave and is in the process of being shredded by fluid
instabilities \citep{Kle94}. 
X-ray imaging contradicts this interpretation, however, 
because the knot is located at the
apex of a large-scale (0.5 degree) indentation in the eastern limb as
traced by very low surface brightness X-ray emission \citep{Gra95,Lev99}.
Thus, the obstacle is certainly large, extending at least 5 pc 
along the line of sight, and the interaction is at an
early stage.

The optical emission is
confined to a $2\arcmin \times 4\arcmin$ region, but 
the apparent insignificance of the southeast knot may belie its importance.
The highest surface brightest optical and X-ray emitting regions of
the Cygnus Loop are associated with the well-developed reflected and
transmitted shocks that form in mature cloud--blast-wave interactions
\citep[e.g.,][]{Hes94}.  The morphology of these bright
regions is notoriously complex because multiple shocks are present
along the same line of sight. Only in a few cases is the geometry
unambiguous, such as the western limb \citep{Lev96}.  In addition, the
development of fluid dynamic instabilities into the non-linear regime
during the late phase of interaction adds to the challenge of
interpreting the more prominent regions.  These difficulties suggest
that focussing on very recent interactions, although
intrinsically fainter, may provide useful insights into the sizes,
shapes, and density contrasts characterizing the clouds in the
vicinity of the Cygnus Loop.

In this work, we use the
the {\it Hubble Space Telescope} (\hst) to examine
the variation in optical line emission on scales of 
$0\farcs1 \equiv 7\times10^{14}$ cm, assuming a distance of 440 pc.  
Clarifying the geometry of the 
emitting regions, we trace the motion of the blast wave in this cloud interaction
and quantify the relevant physical processes.  We present the data in \S \ref{sec:data},
discuss the morphology and physics in \S\S \ref{sec:morph} and \ref{sec:phys}, 
respectively, and summarize our conclusions in \S \ref{sec:concl}.

\includegraphics[width=3.5in]{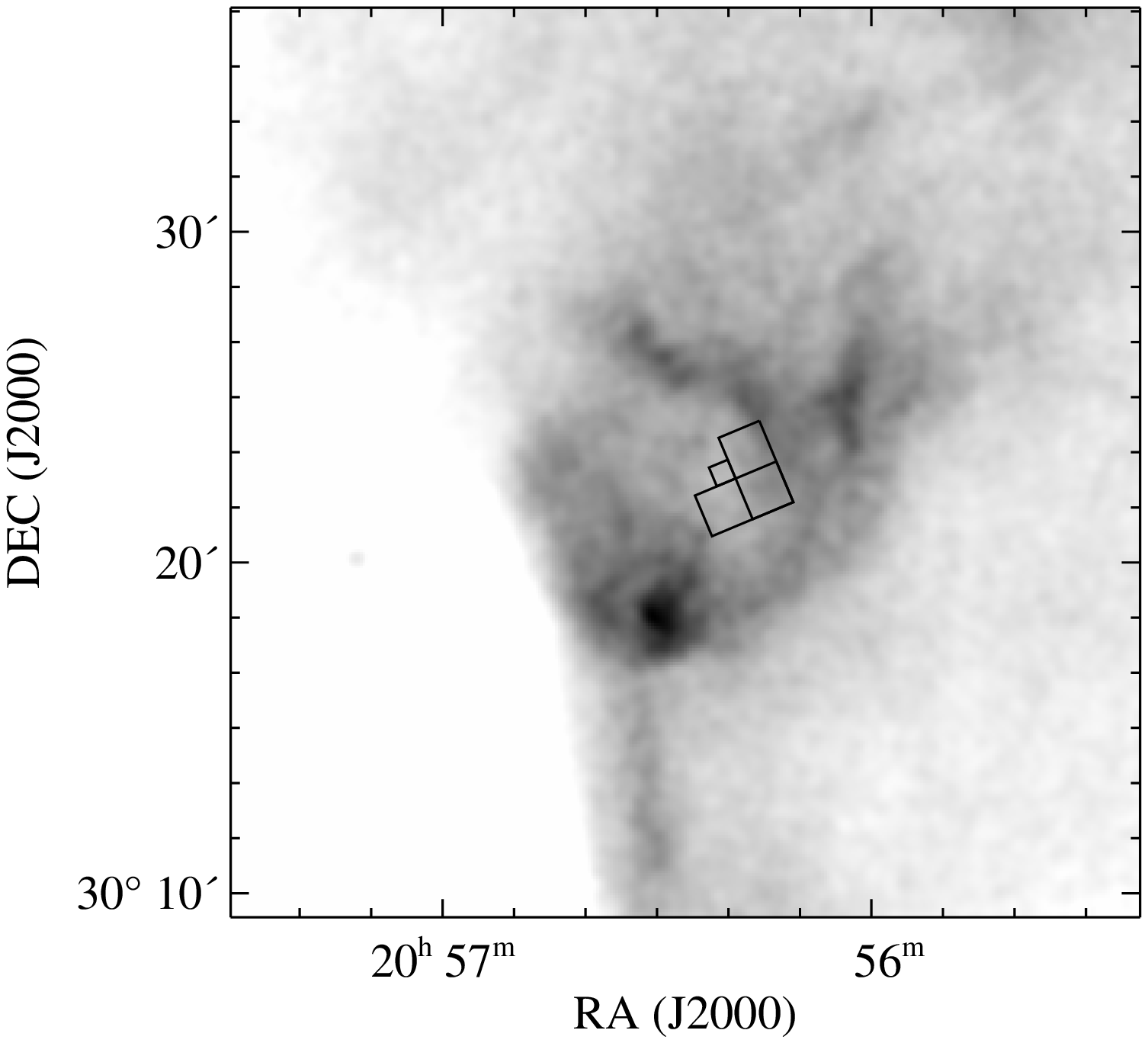} 
\figcaption{\label{fig:xray}
WFPC2 field of view overlaid on 
soft X-ray image of the surrounding field.
The bright optical emission is not coincident with
the associated X-ray enhancement.
}

\section{Observations and Data Reduction\label{sec:data}}
\begin{figure*}[htb]
\centering
\includegraphics[width=6in]{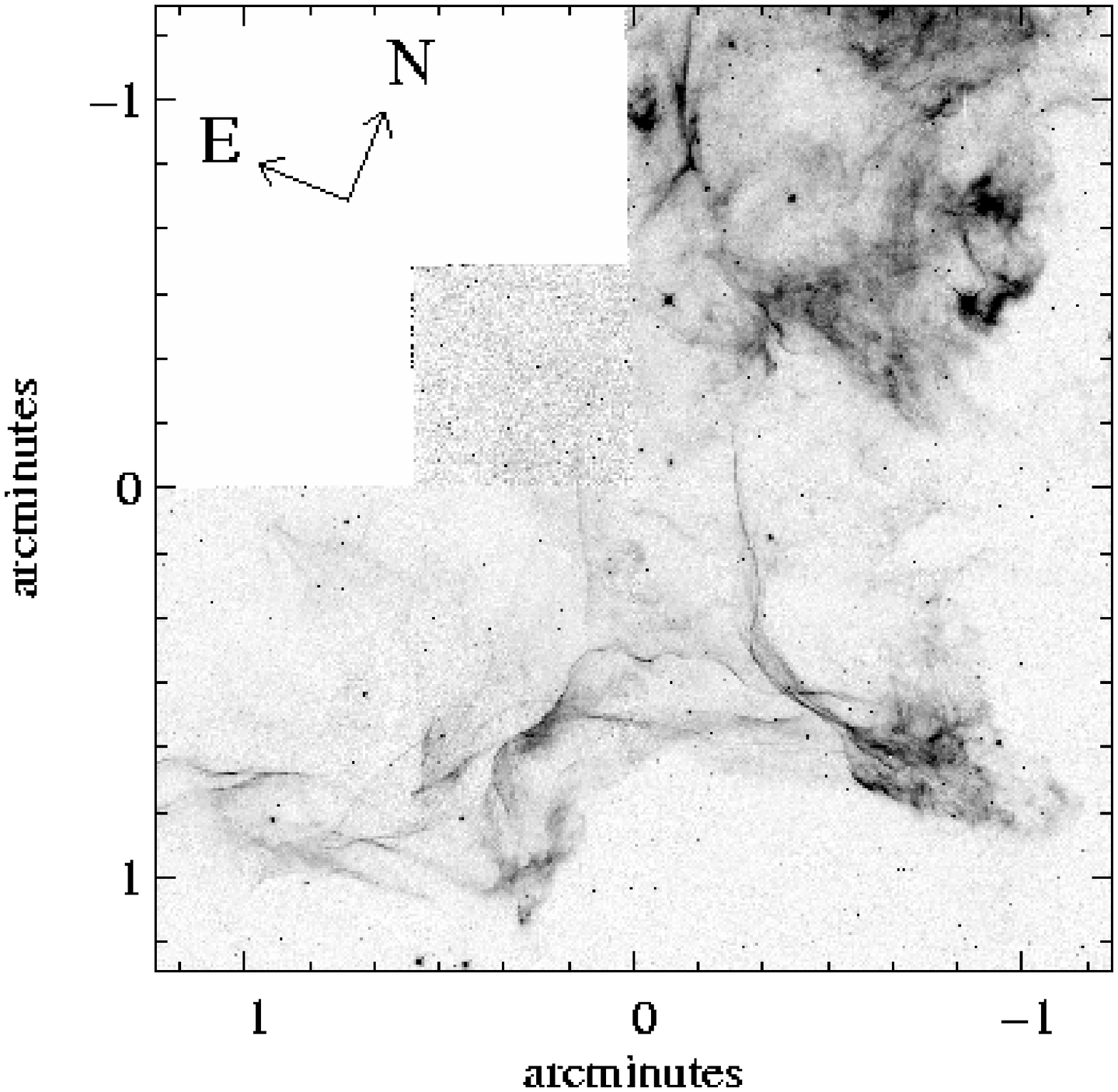} 
\caption{\label{fig:ha}
Southeast knot in the light of H$\alpha$.  The image is scaled
linearly from 0 (white) to $4\times10^{-14} {\rm \,erg\,cm^{-2}\,s^{-1}\,pixel^{-1}}$
(black).
Shocks viewed face-on produce diffuse emission, while 
sharp filaments are characteristic of edge-on shocks.
The upper left quadrant contains the smaller PC1 detector, with WF2 
at the lower left, WF3 at the lower right, and WF4 at the
upper right.
}
\end{figure*}

We obtained the data from the \hst{} archive at the Space Telescope
Science Institute.  
The observations with the Wide-Field Planetary Camera 2 (WFPC2)
were performed 1994 November 25 for Program 5774 
(Principal Investigator J. Hester). 
Figure \ref{fig:xray} illustrates the location and orientation
of these observations on the soft X-ray image 
of the surrounding field obtained with the 
{\it ROSAT} High Resolution Imager \citep{Gra95}.
A total exposure of 2200 s in two frames (to facilitate removal
of cosmic rays) was taken through 
each of three narrow filters.  The F502N filter includes 
\o3{} $\lambda 5007$, F656N includes \ha, and F673N includes
\s2 $\lambda\lambda 6717+6731$.
We employed standard \hst{} pipeline processing, then used the 
IRAF\footnote{IRAF is distributed by the National Optical Astronomy
Observatories, operated by the Association of Universities for
Research in Astronomy, Inc., under cooperative agreement
with the National Science Foundation.}
task {\tt crrej} to remove cosmic rays in the average images.
We assembled the individual detector images into a single mosaic
using IRAF's {\tt wmosaic}.
Figure \ref{fig:ha} contains the \ha{} image,
with the detectors PC1, WF2, WF3, and
WF4 identified counterclockwise, beginning at the upper left.
The $0\farcs046$ pixels of the PC1 correspond to $3\times 10^{14}{\rm \, cm}$,
and the $0\farcs1$ WF pixels are equivalent to 
$7\times10^{14} {\rm \,cm}$.
We apply the flux calibration of \citet{Hol95},
using the 1998 calibration.
The system throughput is 0.104, 0.111, and 0.052 
for \ha, \s2, and \o3, respectively \citep{HST98}. 

\section{Morphology\label{sec:morph}}
\begin{figure*}[htb]
\centering
\includegraphics[width=6in]{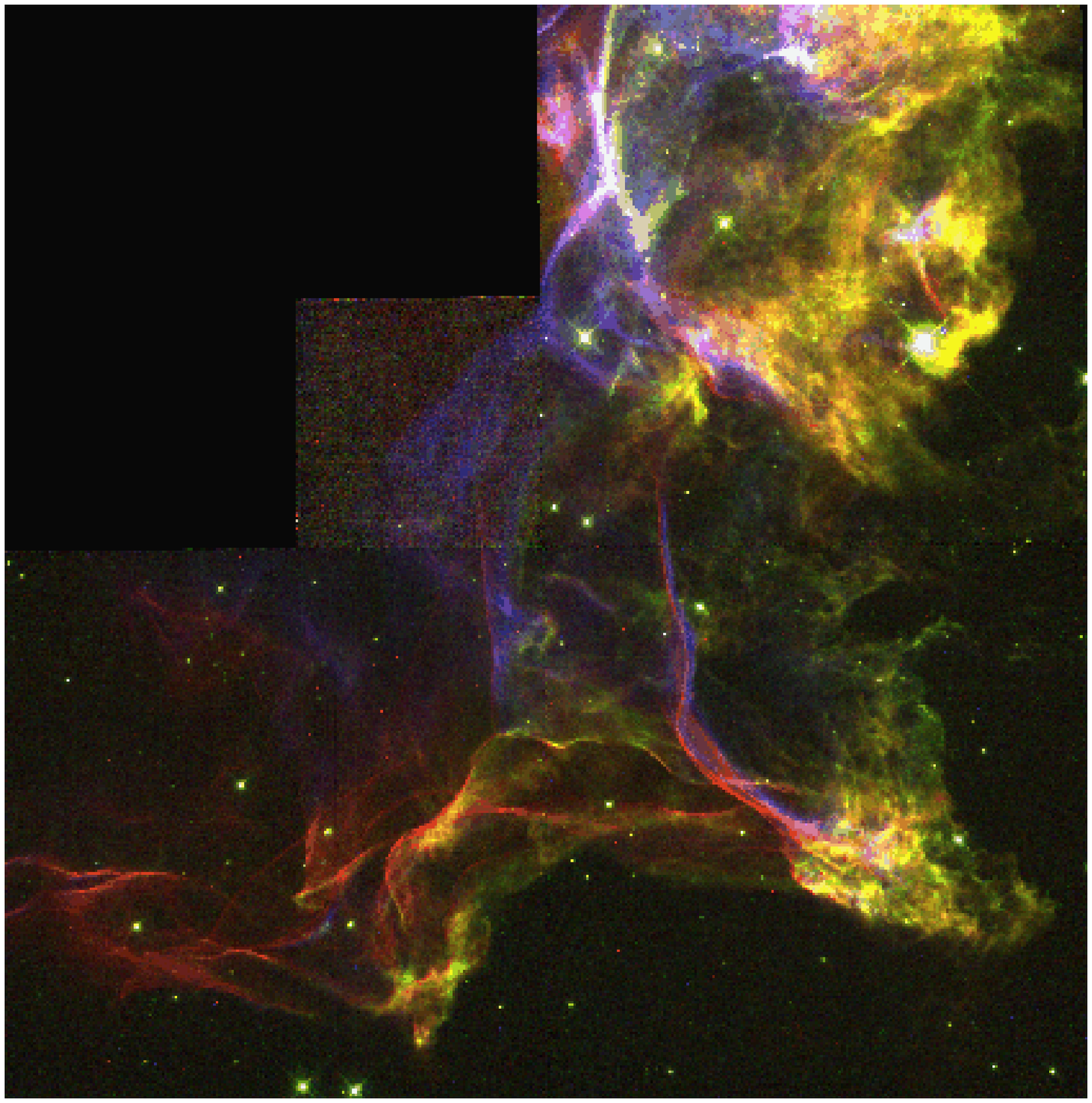}
\caption{\label{fig:color}
False-color image of the southeast knot.  H$\alpha$ emission is mapped in
red, \s2{} in green, and \o3{} in blue.  
The complete radiative cooling zone has developed in
bright yellow regions, with strong H$\alpha$ and \s2.
The bright ``heel'' at the
lower right of the image marks the distinct cloud edge.
Pure red filaments occur
at non-radiative shock fronts.  In the 
examples that have swept up sufficient column density, 
\o3{} emission appears close behind, or is dominant, without
obvious associated H$\alpha$ filaments, in 
regions of incomplete cooling.
}
\end{figure*}

The supernova blast wave moves from west to
east across the field of view and has recently encountered a cloud of
denser-than-average interstellar gas. In the \hst{} field, we observe
the southern section of the interaction, which extends
out of the field of view to the north for another $4\arcmin$.
We combine the three narrow-band data sets in a false-color image
(Figure \ref{fig:color}), with \ha{} mapped in red, \s2{} in green,
and \o3{} in blue.
Magenta thus corresponds to strong \ha{} and \o3,
while yellow shows where \ha{} and \s2{} appear together.
Cyan, which would come from strong coincident \o3{} and \s2\, is almost entirely
absent from the image.

In Figure \ref{fig:color}, color and morphology are strongly correlated.  The shock morphology is
distinctly stratified, 
with
several characteristic types of filaments dominating the composition
of the image.  
Similar to larger-scale observations of the Cygnus Loop at lower
resolution, the two fundamental characteristic
morphologies---sharp filaments and diffuse emission---correspond to
two distinct viewing geometries \citep{Hes87}.  The sharp filaments arise in 
wavy sheets viewed edge-on, through tangencies, and in general these
shocks propagate in the plane of the sky.  This is
the most favorable geometry for unambiguously viewing the stratification
of the post-shock flow as it cools and recombines.
The images therefore 
reveal in turn the Balmer-dominated
and \o3-dominated regions behind the shock front.
In contrast, shocks
viewed face-on produce diffuse emission.  Because larger 
swept-up column densities are required for this diffuse
emission to be detectable, it is more likely to exhibit the characteristics
of a complete radiative shock, namely strong \s2{} emission relative
to \o3{} and \ha.

Toward the north and west the morphology
becomes less filamentary, 
and secondary colors (magenta and yellow), as opposed to primary colors, become more common, with the superposition of multiple cooling stages observed
along the line of sight.
Green and yellow are prevalent in the western sections of the southern
filaments.  This is most apparent at the heel of the emission, 
which delimits the western extent of the cloud in WF3, the lower
right quadrant of the image.  The bright
yellow complex at this location traces strong H$\alpha$
and \s2, which is characteristic of shocks that have swept up a
substantial column ($N_H \gtrsim 10^{18} {\rm \,cm^{-2}}$) 
and formed complete cooling and recombination
zones.
   
The slowest
shocks show up as \s2{} (green) only; occasionally these
regions appear as high surface brightness knots (e.g., at the bottom
edge of WF2, the lower left quadrant). 
In these instances, the primary blast wave is strongly decelerated
in the dense cloud medium.
More common is the extended, faint \s2{} emission, 
which is widespread across WF2 and WF3. 
In general, this
relatively smooth \s2{} emission occurs by itself, unassociated with
\ha{} or other filaments.  This low surface
brightness \s2{} emission is bounded on the western side by a sharp
edge that delineates the current location of the shock
within the cloud.  This edge can be traced from the bright green knots
at the western limit of WF2, then running north and northwest into WF3,
around the bright heel and to the northeast up into the bright complex in 
WF4 at the upper right.
This edge probably represents the original and undisturbed surface of the cloud,
since 
the interaction is recent and has not yet had a significant dynamical effect on
the cloud.
In several locations, we identify the initial development of the
radiative zone, finding \s2{} emission downstream of \ha{} filaments
(near the right center of WF2 and at the bottom of the heel in WF3,
for example).
The \s2{} emission is typically offset behind the shock front by
$0\farcs5$ ($3 \times10^{15}$ cm).  Unlike the sharply-peaked 
\ha{} filaments,
the \s2-emitting region is resolved, 
with flux extending over an arcsecond-scale region of
the sky behind any distinct
portion of the shock front.

The brightest incomplete \o3{} filaments and the \s2-emitting
shocks in the heel region are clearly physically associated. At the
center of the field of view the two main \ha/\o3{} filaments
are part of a segment of blast wave that is propagating to the southeast. Two
tangencies to this surface form the two most prominent incomplete 
\o3{} shocks and their preceding Balmer filaments.  The filament
turns through 45 degrees to form a funnel-like cusp with the heel.
Tracing the upper filament towards the heel, the \o3{} 
emission first merges with the H$\alpha$, producing magenta in the
false-color image.
It then disappears as 
it eventually joins the shocks driven into the western extremities of the
cloud.  Similar morphologies are formed by the blast-wave--cloud
collision farther north in the XA region \citep{Hes86}. 

Emission from shocks with well-developed cooling and recombination
zones comprise the northern section of the interaction and fill the WF4
field. The absence of large-scale filamentary structures implies that
these shocks are more nearly face-on and are lighting up the surface
of the cloud. The sharp filamentary structures in this region have 
\o3{} emission, appearing blue and magenta, and can be connected to the
Balmer-dominated and incomplete shocks farther south.  These less-decelerated shocks
appear in projection against this northern field, 
so the relative east-west position does not correspond to absolute advancement
within the cloud. 

The prominent, sharp-edged, red structures are
Balmer-dominated filaments and appear as a part of a continuous, gently
rippled sheet, as in the extreme southern section of WF2.  These filaments
are due to fast shocks ($v_s > 100 \kms$) that excite H$\alpha$
emission by 
electron collisions in pre-shock gas that
is predominantly neutral and atomic \citep{Che78,Che80}.  
The excitation is confined to in a
narrow zone immediately behind the shock front, and 
the resultant face-on H$\alpha$
surface brightness is low.   Thus, in these
``non-radiative'' shocks, the 
filaments are seen as bright, sharp structures when the shock front is
close to tangency with the line of sight 
\citep[cf.][]{Hes87}. 
Since the gas must be
neutral to produce these filaments, they also mark regions where the
gas is being shocked for the first time, delineating the outer
boundary of the blast wave. Gas-dynamic phenomena in which the gas
is multiply shocked, including reflected shocks, cannot produce 
Balmer-dominated filaments.

The Balmer filaments with no associated \o3{} (blue) emission have
swept up $N_H< 10^{17}\psc$. These shocks has suffered the smallest
deceleration.
While
the general trend in this image is for shocks to propagate from
west (left) to east (right), filaments at skewed angles 
reveal their
direction of propagation when they have swept up
sufficient column for \o3{} ($N_H \gtrsim 10^{17} \psc$) or \s2{} 
($N_H \gtrsim 10^{18} \psc $) emission to be detectable downstream.

At the center of the field are two good examples of patches of
\o3{} emission that are correlated with Balmer-line filaments.  Their
appearance is consistent with a volume of \o3{} emission bounded on
the upstream side by the Balmer-line-delimited shock transition.  The
shock at this location is propagating to the southeast.  Farther 
east in the PC1 is an amorphous region of pure \o3.  Any
associated H$\alpha$ is very faint, suggesting that the direction of
propagation of the blast wave here is more nearly face-on.  The 
\o3-dominated incomplete shocks extend to the top of the image and
are interspersed with and project against more complex and diverse emission
morphologies.

\section{Non-Radiative Filaments and Incomplete Shocks\label{sec:phys}}
In these images, the relationship between the Balmer filaments
and the downstream \o3{} emission of the incomplete shock
distinguishes the physical conditions that are present.  
We examine the filament near the center of 
the WF3 field, which provides a particularly clear example.
At this location, a Balmer-line filament bounds an incomplete
   cooling and recombination zone.
The ratio of \o3{} to \ha{}
surface brightness reaches a plateau of about 6 at a distance
$5 \times10^{15}$ cm behind
the current location of the shock front, which the center of 
the \ha{} emission defines.  This value exceeds the maximum 
of about 2 that can occur in fully radiative shocks,
identifying it unambiguously as an incomplete shock.

\includegraphics[width=3.5in]{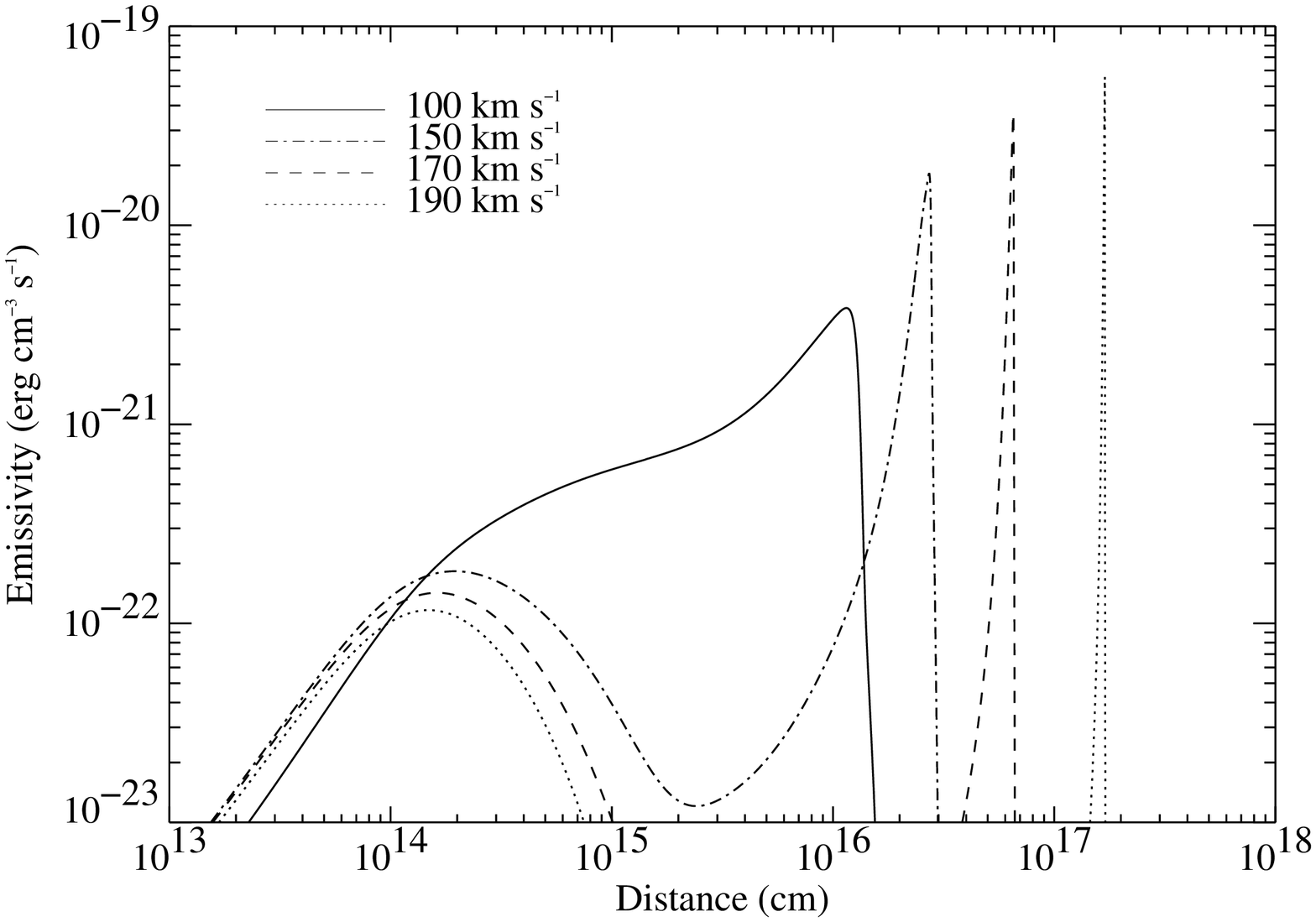}
\figcaption{\label{fig:models}
Model \o3{} line emissivity  as a function
of distance behind shock front for various shock
velocities and $n=1 {\rm \, cm^{-3}}$.
While in the $v_s=100\kms$ case, the profile rises
gradually and reaches a maximum $10^{16}$ cm behind
the shock front, 
in the faster examples, the brightest emission
is sharply peaked farther behind
the shock. 
}
\vskip0.1in

The key
parameter that determines the location and width of the \o3{} 
emission zone 
is swept-up column density behind the shock front.
More exactly, 
the  \o3{} surface brightness profile  constrains the ambient density
and shock velocity.  In general, \o3{} emission rises gradually
and closer to slower shock fronts, while the profile is sharply peaked
and offset farther downstream from fast shocks, as Figure \ref{fig:models}
illustrates. 
The effect of increasing density is to shift the primary \o3{}
peak closer to the shock front.
(In all cases, the initial, smaller peak about $10^{14}$ cm behind the shock
front occurs while oxygen is excited to higher ionization states,
when \o3{} is not the dominant coolant.)

\includegraphics[width=3.5in]{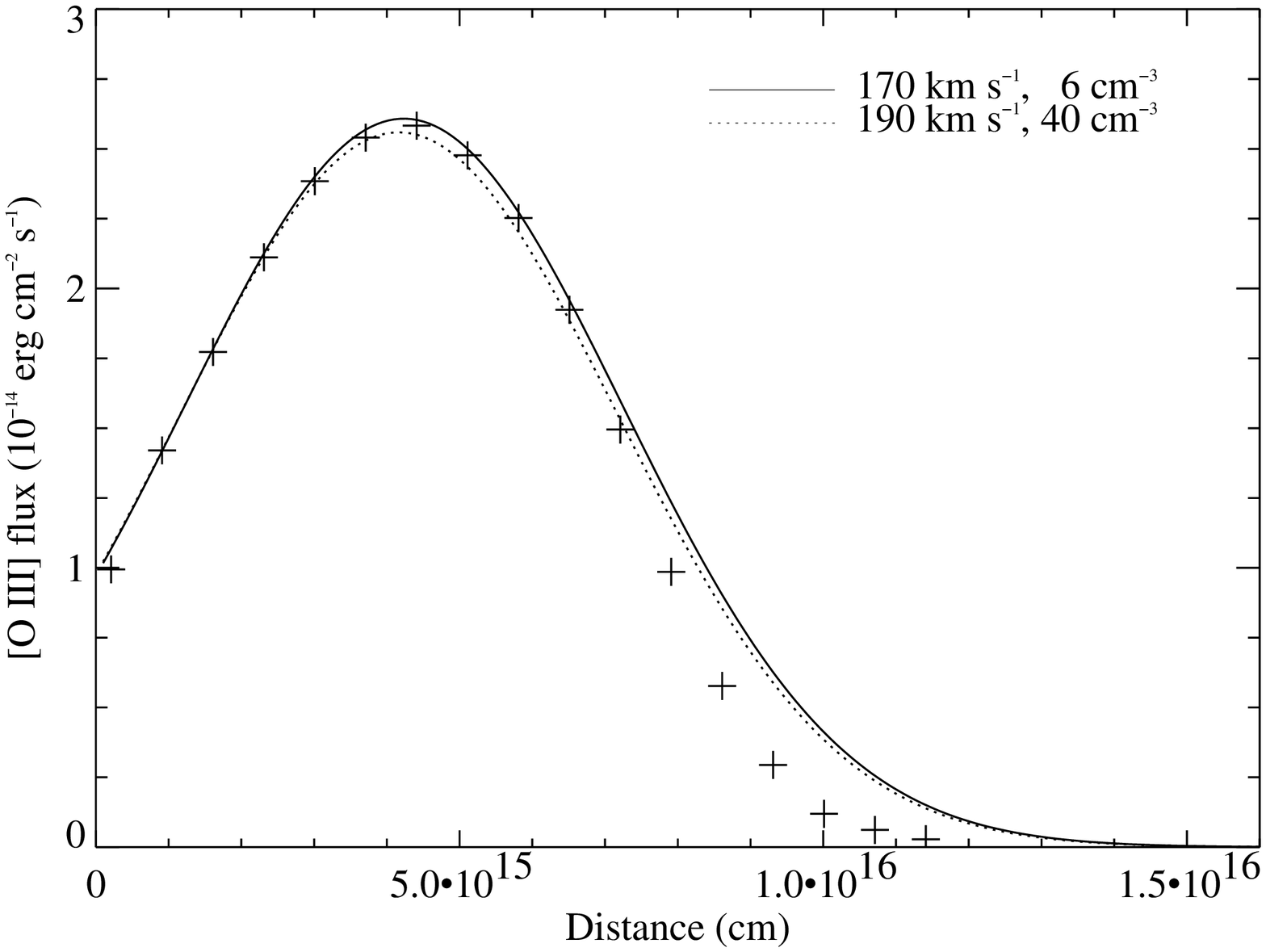}
\figcaption{\label{fig:datamod}
Observed and model 
\o3{} emission profiles.  We find good agreement between the
data ({\it crosses}) and model profiles 
having $v_s = 170$ ({\it solid line}) to 
190 ({\it dotted line}) \kms, 
assuming line of sight depths of $9\times10^{17}$ and
$1 \times 10^{18}$ cm, respectively.
}
\vskip0.1in

We apply updated versions of the models 
described by \citet{Ray79} and \citet{Cox85}
for shock velocities
ranging from 100 to 190 \kms, 
convolving these models with the 
instrumental profile measured from point sources in the field of view.
The Balmer filament fixes the location of the shock transition,
and the free parameters are shock velocity, pre-shock density, and
depth of the surface  along the line of sight.
While the models include magnetic fields, in the hot, post-shock gas
these data trace, 
thermal pressure dominates, and the effect of magnetic pressure
is negligible.
We find good agreement with both $v_s = 170$ and $190 \kms$,
with density $n = 15$ and $40 {\rm \, cm^{-3}}$  respectively, in these two
cases (Figure \ref{fig:datamod}).
We cannot significantly distinguish between these
fast shocks  because
the instrumental resolution dominates the emission profile,
but we prefer $v_s = 170 \kms$ because 
the location of peak emission of
this model better matches the observations.
The residuals between the model and data at distances
around $10^{16}$ cm are likely the result of poor subtraction
of the local background where the flux is extremely low.
Thus, we adopt the model parameters $v_s = 170 \kms$ and
$n = 15 {\rm \, cm^{-3}}$.
In calculating the total flux, we initially assume that the
geometry of the emission region is a thin sheet that
extends $3\times10^{17}$ cm in the line of sight,
which is
the observed extent of the filament across the image.
Comparison of the observed and predicted intensities
implies that the line of sight depth of the filament
is in fact factors of 3 to 4 times its extent in
the plane of the sky.
This is a natural result because our selection bias
favors bright filaments, which tend to be those with
the greatest extent along the line of sight. 
The inferred velocity is robust against uncertainties 
in the assumed distance to the Cygnus Loop, while the
derived densities are inversely proportional to distance.

For comparison, we compute the pressure from the X-ray observations
of the same region.  The surface brightness measured with the
{\it ROSAT} High Resolution Imager 
corresponds to an emission measure of $70 {\rm \,cm^{-5}\,pc}$.
Here we assume a temperature $T=2\times10^6$ K and solar abundances
with two caveats:
the temperature in this small region may be
somewhat different than the average temperature, which is
weighted by the brightest X-ray emission; and gas-phase depletion
and subsequent grain destruction may alter the abundances \citep{Van94}.
Using the line of sight depth $10^{18}$ cm determined above
and  a 20\% contribution of metals to the electron
density, $n_e$, 
we require $n_e= 16 {\rm \,cm^{-3}}$.  Thus, pressure 
$P=8.7 \times10^{-9} {\rm \, dyne\,cm^{-2}}$, which is comparable
to the ram pressure that drives the cloud shock, based on 
the best-fitting parameters above.

These Balmer filaments and their associated incomplete shocks
are distinct from those observed around the perimeter of
the Cygnus Loop.  In those cases, the filaments define
the nearly-circular undisturbed blast wave and are continuous
over scales of $40\arcmin$ \citep{Lev98}.  The corresponding
shocks 
propagate through a much less dense
medium than the southeast cloud provides.
\citet{San00}, for example, find $n = 2$--$4 {\rm \, cm^{-3}}$ and
$v_s \sim 170 \kms$ in a filament on the northeast periphery,
based on ultraviolet line fluxes and intensity ratios.
Assuming the same initial blast wave properties
at the northeast limb and in the southeast knot,
we conclude that the latter must be a younger interaction.

The overpressure that has developed behind the southeast shock
drives it strongly into the cloud at a pressure
exceeding that of the primary
blast wave,  as measured in other cloud encounters of the 
Cygnus Loop \citep{Ray88,Hes94}.
The maximum overpressure of a cloud interaction
is a factor of 3 once steady flow is established \citep{McK75},  
or up to a factor of 6 when the blast wave encounters a plane of
material \citep{Spi82}, which is more similar to this
very early stage of interaction with a large cloud.
The greater overpressure of the southeast knot 
is indicated not only in comparison
with the northeastern filament, but also 
(and more significantly) when the average blast
wave pressure, $P_{BW} \approx 5\times10^{-10} {\rm \, dyne\, cm^{-2}}$, 
derived from global X-ray data \citep{Ku84} is considered.
The large overpressure at the southeast knot therefore indicates
highly non-steady flow.
A shock having $v_s = 170 \kms$ is unstable
unless the transverse magnetic field $B \gtrsim 10 \mu$G,
but because the timescale for the development of secondary
shocks is long (on the order of  $10^4$ years), they do not appear yet
in this case \citep{Inn92}.
Thus, the transient nature of the current conditions 
is expected. 
Only later will 
a slower, large-scale, coherent
shock arise in a more developed interaction,
similar to the western edge of the Cygnus Loop, for example.

Multiple shocks along the line of sight confuse the 
\o3{} profile in several other  
locations where Balmer emission bounds a region of 
bright \o3, so we cannot directly 
compare with the models, but we can characterize them qualitatively. 
We identify 
the Balmer filament at the center of the mosaicked field 
as a region of lower density, because the \o3{} emission is 
broader and offset farther downstream. 
Immediately behind the filament at the west-center of the WF2 field,  
the fully-radiative signature of \s2{} implies that this 
is a higher density region.  The emission
at the extreme southeast of the WF2 field consists exclusively of
\ha, which suggests that the shock propagates through lower density here,
perhaps in the extended envelope of the cloud.  
Furthermore, this region includes not only the sharp filaments
that characterize the edge-on view, but also more diffuse
emission where the shock surface is viewed at oblique
angles.  Thus, this region must be {\em intrinsically} bright.
The emissivity depends linearly on $v_s$ and $n$ \citep{Ray91},
so the shock velocity is higher here.

\section{Conclusions\label{sec:concl}}
The \hst{} images of the interaction between the Cygnus Loop
blast wave and an interstellar cloud reveal emission
variations 
on the smallest measurable scales ($3\times 10^{14}$ cm).
The blast wave cannot be identified as a single, uniform
entity, but is broken into a complex of interacting
shock fronts as it encounters the obstacle.  We identify the characteristic
morphology of sharp filaments, where a shock front is viewed
edge-on, and diffuse emission, where the view is face-on,
on the sub-arcsecond scales that WFPC2 probes.  The \o3{}
profile immediately behind the shock front that Balmer-dominated
filaments define reveals relatively fast shocks ($v_s \approx 170 \kms$)
in the high-density ($n \approx 15 {\rm \,cm^{-3}}$) cloud medium.
Balmer filaments without associated \o3{} or \s2{} emission
arise in slightly lower-density regions behind faster
shocks.  Because this emission is intrinsically bright, we
detect the diffuse component, which is viewed obliquely, 
as well as the more common 
sharp filaments where the shock front is viewed edge-on.
Exhibiting extensive networks of 
non-radiative shocks, 
the southeast knot must represent the early stage of interaction
between the cloud and blast wave.
This example thus 
illustrates that complex shock propagation and emission morphology
occur before the onset of  
instabilities that destroy clouds completely.

\begin{acknowledgements} 
We thank Ravi Sankrit for computing the shock model emission profiles.
Support for this work was provided by NASA through grant number
AR-08005
from the Space Telescope Science Institute,
which is operated by AURA, Inc., under NASA contract NAS 5-26555.
\end{acknowledgements}

\end{document}